\documentclass[usenatbib]{mn2e}

\usepackage{graphicx}
\usepackage[space]{grffile}
\usepackage{latexsym}
\usepackage{textcomp}
\usepackage{longtable}
\usepackage{multirow,booktabs}
\usepackage{amsfonts,amsmath,amssymb}
\usepackage{natbib}
\usepackage{url}
\usepackage{hyperref}
\hypersetup{colorlinks=false,pdfborder={0 0 0}}
\newif\iflatexml\latexmlfalse
\usepackage[utf8]{inputenc}
\usepackage[english]{babel}

\title[Light Echoes from Linear Filaments]{Light Echoes from Linear Filaments}

 \author[Nemiroff \& Zhong]{
Robert J. Nemiroff$^{1}$\thanks{E-mail: nemiroff@mtu.edu}
and Qi Zhong$^{1}$
\\
$^{1}$Michigan Technological University, Dept. of Physics, 1400 Townsend Drive, Houghton, MI 49931 \\
}

\begin{document}



\maketitle

\label{firstpage}

\begin{abstract}
When a flash of light from a star overtakes a straight linear filament of gas or dust and is seen later by an observer, a pattern of perceived illumination occurs that encodes information about the distance to the flash, the distance to illumination fronts on the filament, and the orientation of the filament. To help decode this information, geometric considerations of light echoes from such filaments are considered. A distinction is made between real spots, which occur unambiguously on a filament, and perceived spot echoes, which are seen by observers and may appear differently to separated observers. For context, a series of critical points are defined on a hypothetically infinite filament. Real spot pair creation events will only occur on an infinite filament at the closest distance to the flash, while perceived spot pair events will only occur when the radial speed component toward the observer of a real spot crosses the speed of light. If seen, a perceived spot pair creation event could provide unique information toward decoding distance and orientation information of the flash and the filament. On filament segments, typically only one of these perceived spots will be seen. Geometries where a perceived spot appears to move with an angular component toward the flash are shown possible. Echo and source distance determinations for filaments that pass between the observer and flash are considered. Hypothetical examples are given for Merope variably illuminating IC 349, and Rigel creating perceived spots on IC 2118.
\end{abstract}


\section{Introduction} 
A flash of light from an object seen later in scattered light, or light absorbed and later re-emitted even in another wavelength band, is known as a light echo. The first recorded case of light echoes was observed as expanding rings around Nova Persei 1901 by \citet{Ritchey_1901} and recognized as light echoes by \citet{1902PA.....10..124K}. The mathematical theory of light echoes was first expounded upon by \citet{1939AnAp....2..271C}. \citet{1940RvMP...12...66Z} suggested that unrecorded supernovae could be studied by later recorded light echoes. That ultraviolet light could heat dust which would then glow in infrared light-echoes was discussed first in detail by \citet{1983ApJ...274..175D}. The theory behind recovering information from light echoes around supernova was given by \citet{Chevalier_1986} and applied to SN 1987A, notably by \citet{1987ApJ...323L..47S} and \citet{1988ApJ...333L..51C}. Works discussing important cases of of light echoes from SN 1987A include \citet{2005ApJ...627..888S} and \citet{2005ApJS..159...60S}.  A good review of light echoes from Type Ia supernova is given by \citet{2005MNRAS.357.1161P}. Additionally, \citet{2003AJ....126.1939S} focused on echoes in the vicinity of variable stars and novas. Since then, light echoes have been found and analyzed around many flashing objects -- for a recent review see, for example, \citet{2012PASA...29..466R}.

Many light echoes to date are recorded as rings -- or ring fragments -- expanding angularly away from a highly variable source such as a nova or supernova. A good discussion of how these rings appear to expand is given by \citet{2004A&A...414..223T}, in particular using V838 Mon as an example. Light echoes confined to a small section of the scattering surface or filament are here referred to as "spots". Relative spot motions between two infrared observations indicate spot motion on filaments of Cas A \citep{2005Sci...308.1604K}. The possibility that light echoes could be found scattered from linear filaments is mentioned by \citet{2011ApJ...732....2R}. 
	
Recently \citet{2015PASA...32....1N} suggested that scattering surfaces near sweeping beams of light can undergo bright spot echo pair-creation events that might be discoverable in astronomical settings. Soon thereafter, such an image pair event was created and observed in a lab \cite{2016SciA....2E1691C}. It has been realized that sonic booms are a common sound analog of this phenomenon \cite{nemiroff_2016_49871}. A major impetus for this work was the realization that such pair events could also occur when the light source is a flash.

This work focuses on light echoes from compact flashes by straight linear filaments. Such filaments are simple approximations to many real filaments that are observed to occur in astronomical settings. The relatively straightforward response of light echoes in this scenario allows for insight to be gained into actual filaments that are nearly straight or have straight components as well as scattering from more complex filaments. 

This paper is structured as follows. Section 2 will review background concepts and define terms that make it convenient to understand and describe how light echoes may occur and appear from flashes on straight linear filaments. Section 3 will explore the general properties of spot motions, while Section 4 will explore how observations can yield the distance and orientation of a filament. Section 5 will explore scattering from filaments that cross between the observer and the flash. Section 6 will briefly touch on relative brightness issues, while Section 7 will discuss filaments in the context of two real-sky examples. Section 8 will summarize results and give conclusions.

\section{Background Concepts}

The two most prominent locations that will be considered will be that of the flash and that of the observer. In general, for simplicity, these two locations will be here considered effectively at rest with respect to each other and at rest with respect to all scattering media. The flash-observer distance will be labeled $d$.

Two important times are the time that the flash occurs, defined as $t = 0$, and the time the flash is witnessed by the observer, defined as $t_f = d/c$ and $\tau = 0$. Here $c$ is the speed of light. The time that the observer witnesses the flash scatter from a given filament element will be designated generally as $t$, which can be defined relative to the time of the flash as $\tau = t - t_f$.

The filaments considered here will be assumed to be one-dimensional straight lines. For brevity, they will be referred to here just as "filaments", and typically assumed to be composed of common interstellar materials like gas and dust. The cross section of each filament will be considered to be uniform along its length and small in size compared to the observer-flash distance $d$ and the length of the filament segment. 

A distinction will be made between "real" spots of illumination and "perceived" spots. A real illumination spot actually occurs on the scattering object, whereas a perceived spot echo is what appears illuminated to an observer. The difference between real and perceived illumination fronts is more than semantic. For example, picture an annular ring that is momentarily illuminated by a central flash and seen edge-on by a distant observer. The entire ring undergoes a real illumination, all at once, by the flash. A distant observer, however, will perceive something quite different, with the part of the ring nearest the observer appearing illuminated well before the furthest part of the ring. In this case, there would be no real pair creation event at all, but the observer would perceive a spot pair creation event at the nearest point followed by a spot pair annihilation event at the furthest point. In general, real illumination patterns are unique, while different observers will see different perceived illumination patterns on scattering media.

\subsection{Real Spot Positions}

\begin{figure}
\begin{center}
\includegraphics[width=0.7\columnwidth]{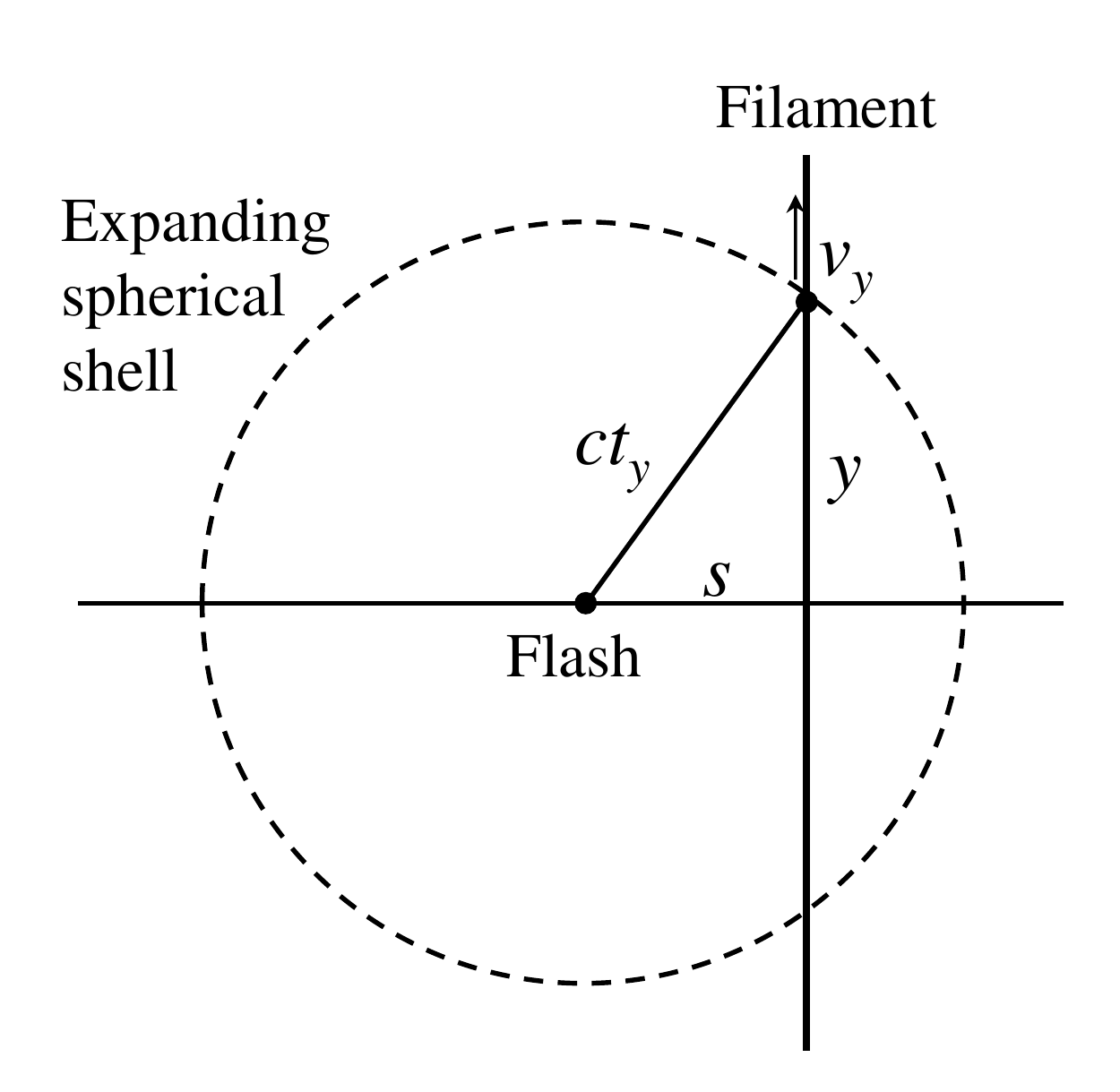}
\caption{{\label{fig:LE}: The expanding spherical shell and the velocity of real light echoes. The solid vertical line represents the linear filament.%
}}
\end{center}
\end{figure}

When a flash occurs, the leading edge is considered to be an expanding sphere of light centered on the flash. Given a flash of finite duration, this leading sphere is followed by a trailing sphere indicating the end of the flash, together creating a spherical shell of light expanding at the speed of light. The thickness of this expanding light shell will be considered small compared to $d$ and the length of any considered filament segment. The resulting small segment of illumination on the filament will be called a "real spot". The geometry is depicted in Figure \ref{fig:LE}.

Conceptually, real spots may be considered the spots that really exist on the filament, irrespective of possible subsequent measurements by separated observers. One might consider a series of time-recording light-sensitive devices distributed over every section on the filament, each device recording if and when it detects light directly from the flash. The positions and motions of real spots might be computable by distant observers given enough information. If so, since these real positions and motions are unique, they are not observer dependent and so if different observers can reconstruct real spot positions and motions, those computations should agree. 

In general, after a flash, the expanding spherical shell of light will first illuminate an infinite straight filament at a single point. This point is the closest to the flash and also the only location on the filament that is tangent to the expanding sphere of light. Therefore, it is here designated as the {\it spherical tangent point}. The geometry is depicted on the left of Figure \ref{fig:SphericalAndEllipsoidal}. After initial contact, two real illumination spots will move on the filament. Because there are two real spots, the initial contact will be referred to as a {\it real spot pair creation event}. 

\subsection{Perceived Spot Directions}

To help describe the spot motions perceived by an observer, several key angular directions on the observer's sky will be defined. All directions will be measured from the flash direction and parameterized by the variable $\theta$. Therefore, by definition, the {\it flash direction} itself will have $\theta = 0$. Related is the {\it anti-flash direction}, the location on the observer's sky 180 degrees around from the flash, delineated by $\theta = \pi$. Although sometimes corresponding to points, the role of perceived events on the observer's sky will be highlighted by referring to them as occurring in specific directions. 

Although infinite filaments do not occur in nature, it is conceptually useful to understand critical spot positions in the greatest theoretical context first, and then focus on subsets that are practically observable. Toward this goal, next defined are the two directions on the observer's sky where the infinite straight filament appears to end: the {\it radiant directions}. Like an infinite straight pole, two unique angular terminal directions on the observer's sky will exist for each filament. Unless oriented perpendicular to the observer-flash axis, one of these two radiant directions will be closer to the angular location of the flash than the other. Therefore, since the two oppositely moving perceived spots are pre-destined to asymptote at these end directions, one perceived spot will approach a radiant direction closer to the flash than the other. 

The location on an infinite filament that is angularly closest to direction of the flash will be dubbed the {\it perigee} direction. This may be a radiant direction of the filament or a direction on the length of the filament. Similarly, the location on the length of an infinite filament that is angularly furthest from the direction of the flash will be dubbed the {\it apogee} direction, which also may be either a radiant direction or on the body of the filament. Note that if a filament has an angular perigee (apogee) direction along its length, then its angular apogee (perigee) direction will coincide with one of the radiant directions.

The locus of points that take time $\tau$ to be recorded by the observer after the perceived time of the flash is an ellipsoid with the flash at one focus and the observer at the other, here called the {\it Time Delay Ellipsoid} (TDE). An example analysis that used TDEs to analyze light echoes was \citet{2008ApJ...685..976D}. At $\tau = 0$ this ellipsoid starts out as a line connecting the flash to the observer. As time progresses, $\tau$ increases and the ellipsoid inflates. The ellipsoid shape is more precisely a prolate spheroid ("cigar shaped") with its major axis on the line connecting the flash to the observer, and its two identical minor axes perpendicular to the major axis. In practice, since most observed astronomical flashes lie between hundreds and billions of light years distant from Earth, and observation durations are typically limited by human lives to be on the time scale of years, then typical TDEs considered here will have a long and thin shape like single uncooked strands of spaghetti.

\begin{figure}
\begin{center}
\includegraphics[width=0.7\columnwidth]{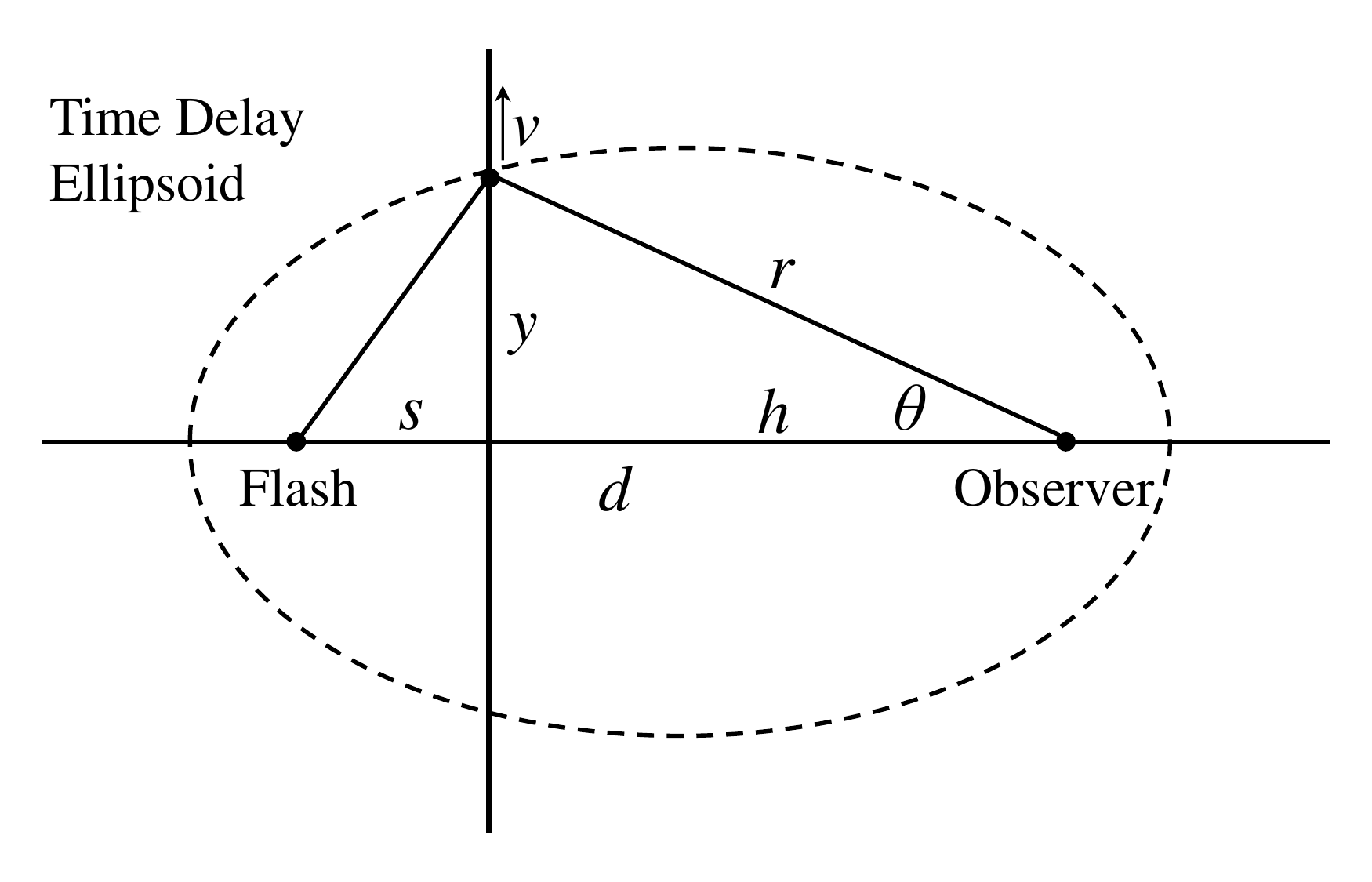}
\caption{{\label{fig:TDE} : A cross section of the time delay ellipsoid (TDE) that includes the flash and the observer. As in Figure 1, the solid vertical line represents the linear filament. In the depicted case, the linear filament lies along -- and is perpendicular to -- the observer's line of sight to the flash.%
}}
\end{center}
\end{figure}

A "perceived spot" of illumination is a small section of a filament that is observed to be illuminated by a previous flash. When the perceived spot is observed to be at angular separation $\theta$ from the flash direction, the distance $r$ from the observer to the perceived spot is 
\begin{equation} \label{PointDistance}
 r = { {c^2 \tau^2  + 2 d c \tau} \over
           {2 c \tau + 2 d - 2 d \cos \theta} } .
\end{equation}
Eq. (\ref{PointDistance}) is derived from the equation of an ellipse in polar coordinates when the origin is taken to be a focal point \citep{2015Wikipedia...DEC...16}. Eq. (\ref{PointDistance}) shows that, given $d$, observing $\tau$ and $\theta$ for a perceived spot is sufficient recover $r$ and so uniquely locate it in three-dimensional space. A perceived spot is located at a real point of illumination at time $r/c$ before observation. The geometry is depicted in Figure \ref{fig:TDE}. 

The expanding time delay ellipsoid (TDE) will first appear to illuminate the infinite filament in a direction here designated as the {\it ellipsoidal tangent direction}. This direction is toward the first point on the filament that is intersected by the expanding TDE, and the only direction where the expanding TDE is tangent to the filament. Given a unique observer, a unique linear filament can only have a single ellipsoidal tangent direction. Different observers, however, will in general identify different directions on the same filament as the ellipsoidal tangent direction. After the apparent first illumination, two perceived spots of illumination will appear to an observer to move away from each other towards opposite ends of the filament. The initial event will be referred to here as a {\it perceived spot pair creation event}. Note that, in general, the directions toward the real and perceived spot pair creation events will be different.  

\begin{figure}
\begin{center}
\includegraphics[width=0.7\columnwidth]{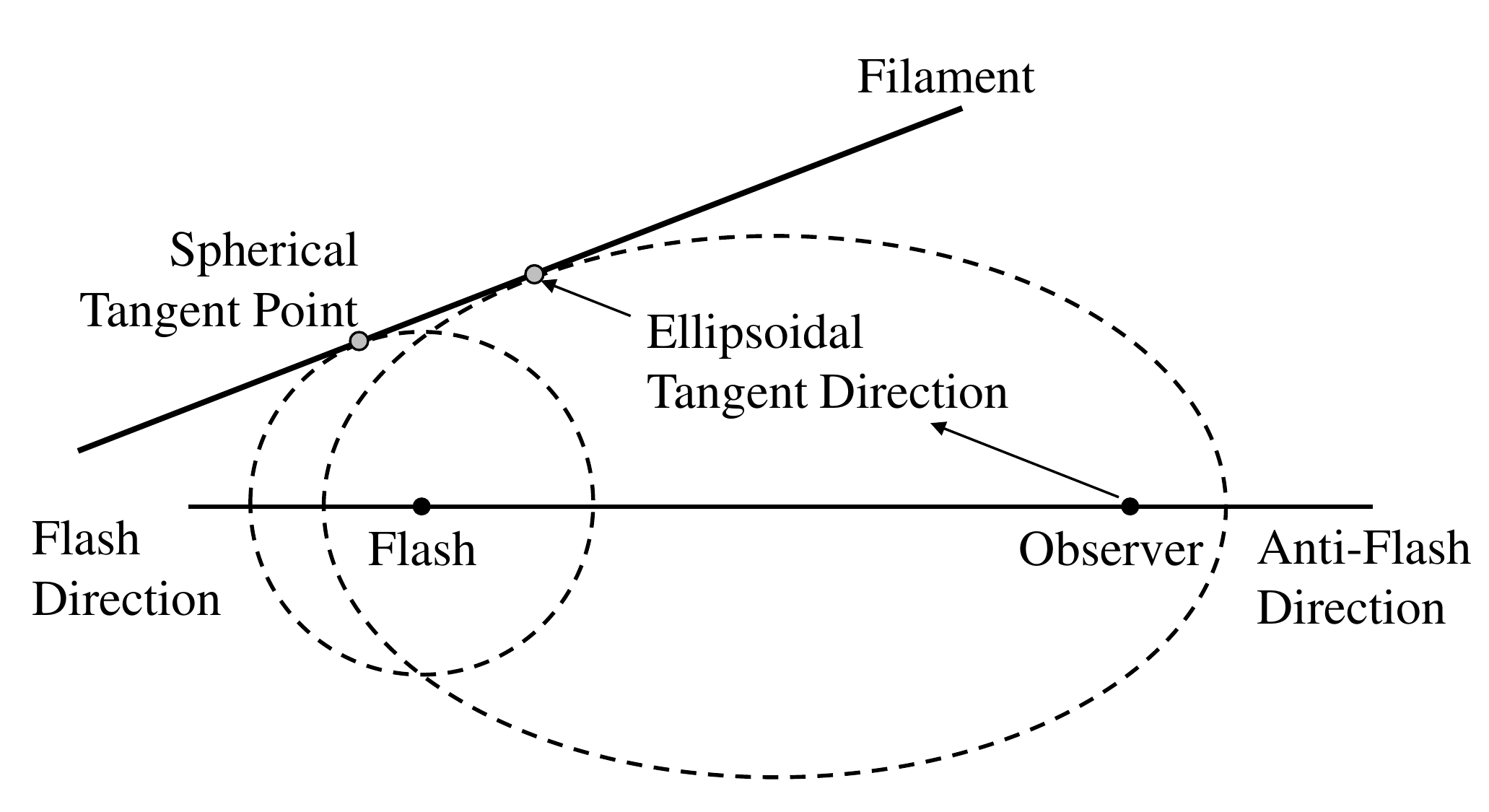}
\caption{{\label{fig:SphericalAndEllipsoidal}: The locations of several points and directions with regard to the flash and the observer.%
}}
\end{center}
\end{figure}

The spherical tangent point and the ellipsoidal tangent direction are depicted in Figure \ref{fig:SphericalAndEllipsoidal}. To be clear, all observers will agree on the location of the unique spherical tangent point in space, which is the point where the real spot pair creation event occurs. However, different observers will note different directions toward their own ellipsoidal tangent directions, the directions toward which each observer would perceive a spot pair creation event to occur.

\section{Spot Motions}

\subsection{Real Spot Motions}

Real spot motions refer to the motions of real spots on filaments irrespective of how they are observed. Understanding real spot motions can, in theory, be agreed upon by all observers, given enough information. Solving for real spot positions and motions may -- or may not -- be useful as an intermediate step to computing observer perceived spot motions. 

How fast do the real spots move? Assume that the flash occurs at a distance $s$ from the spherical tangent point, meaning that $s$ denotes the closest distance between the flash and filament. Since photons move out from the flash at speed $c$, the time that the flash first strikes the filament is $t_1 = s / c$. Take $t_y$ as the time it takes for light from the flash to impact position $y$ along the filament, with $y=0$ being the spherical tangent point. From Figure \ref{fig:LE} it is clear that $y^2 + s^2 = (c t_{y})^2$ so that $ct_y = \sqrt{ y^2+s^2}$. Then the speed of the real illumination front along the filament is
\begin{equation} \label{SpeedActualGeneral}
v_y = {dy \over d t_y} = {\sqrt{y^2+s^2} \over y}c  .
\end{equation}

Note that when $y = 0$ then $v_y$ is formally infinite. Real spot speed drops monotonically with increasing values of $y$, only approaching $c$ far from the spherical tangent point. As indicated in a previous work, the speed of any real illumination front from any flash across any surface is always superluminal \cite{2016PhyEd..51d3005N}.  Therefore, the speed of real spot motions here is seen to conform to this superluminality law. Since the geometry is symmetric about the $y = 0$ origin of the filament, the other real spot created at the same time will have the same magnitude of speed but in the opposite direction, also being always superluminal. 

\subsection{Perceived Spot Motions}

Although perceived spot motions are by definition angular, defining them theoretically is inherently more complicated than for real spot motions because the observer - flash axis forms a line that is generally different from the line that defines the filament. Therefore, since in general two lines do not define a plane, the perceived one-dimensional angular path of perceived spots, in an observer-centered coordinate system,  must be done utilizing all three spatial dimensions. Nevertheless, the perceived spots are constrained to move along the angular line made by the filament, and perceived spot motions have conceptual similarities with real spot motions. For example, there will be a single (real and perceived) spot creation event perceived along the direction of the linear filament, with each (real and perceived) spot from the pair perceived to be moving along the filament toward opposing ends of the filament. 

There are several different routes to computing the directions and motions of the perceived spots on the filament, each of which converges to common mathematics. One conceptual path is to compute the intersections of the expanding TDE with the linear filament. Another conceptual route is to keep track of the time of flight of a photon from the flash to the observer as a function of location on the filament, and find times of minimum flight. Unfortunately, besides general technical descriptions, there appears to be no simple closed-form general mathematical solutions. Some specific solutions that occur when the filament intersects the observer-flash axis are given below. 

How does an observer perceive spots to move? Here infinite linear filaments will again be sometimes be considered to demonstrate perceived spot motions in the greatest theoretical context. A general scenario starts with a flash occurring at $\theta = 0$. For a time afterward, the filament appears completely unilluminated. Suddenly, a perceived spot pair creation event appears toward the infinite filament in the ellipsoidal tangent direction. After creation, one perceived spot moves along the filament initially in the direction of the perigee, while the other moves toward the apogee. Oddly, one perceived spot will cross the direction toward the point where the real pair event occurred -- the spherical tangent point -- without anything unusual appearing to happen. After that, both perceived spots move toward their respective radiant directions, 180 degrees apart on the observer's sky. For filaments of finite length, the creation and motion of perceived spots are the same but only the parts of the filament that actually exist can be seen illuminated. 

It is tempting to assume that the perceived spot creation event occurs in the direction of the real spot creation event, but that is not usually the case. This is because an illumination front with a real speed toward the observer that is faster than $c$ actually appears to that observer to be moving away -- because the real front outpaces the images of the front. Therefore, perceived spot pairs are created only when the radial speed of a real spot toward the observer drops from superluminal to subluminal \cite{2015PASA...32....1N}.  

Oddly, this counter-intuitive incongruence can be used to an observer's advantage. Given a perceived spot pair creation event in a specific direction, the observer knows that the radial velocity of the real spot in this direction must be precisely $c$. In contrast, were a single spot observed, there is degeneracy between angular speed, real speed, and distance. For a perceived spot pair creation event, however, this degeneracy partly resolved as there is only one real speed allowed. Although this resolution can contribute unique information about the distance or orientation of the filament in theory, full mathematical solutions of these may be quite complicated in practice. In general though, given that the direction of motion of a real spot on a linear surface at position $y$ makes an angle $\phi$ to the observer, then $v_y \cos{\phi} =-c$ so that
 \begin{equation}
y_{pair} = c t_y \cos{\phi} .
\end{equation}

\subsection{Perceived angular motion toward the flash}

Geometrically, it is clear from inspection that the expanding TDE cannot first intersect an infinite filament toward a radiant direction. Nothing in the geometry demands that the ellipsoidal tangent direction points toward the perigee direction, and in general those directions will not coincide. Assuming that they do not coincide, and since the two perceived spots move in opposite directions, one perceived spot must appear to move toward the perigee. A result of this is perhaps surprisingly: one of the perceived spots will appear to the observer to initially move on the sky with an angular component {\it toward} the direction of the flash. To the best of our knowledge, such motion has never been seen nor discussed in an astronomical setting. It is mentioned here partly for its novelty and partly to alert observers that apparent motion of a bright patch toward a flash does not necessarily rule out an echo interpretation. In fact, although not described in detail, such motion has been recorded in a lab on Earth \cite{Velten:2013:FCV:2461912.2461928}. Additionally, a video uploaded to YouTube ({\url https://www.youtube.com/watch?v=c8XEVT8URoY}) by the Spitzer Space Telescope group in 2008 shows, coincidentally, some of the effects described here, including, in one case, an infrared light echo actually moving angularly toward of the flash. 

Of course, if the part of the filament that would have shown perceived angular motion toward the flash does not exist, then such motion cannot be perceived by the observer. As will be demonstrated in examples below, it may also occur that the only part of a linear filament that exists is a part that displays one of the image pair. In these cases, only one spot will be perceived to exist and move to the observer.

\section{Filament Orientation and Distance}

Since infinite linear filaments are unrealistic, as are filaments even of length scales comparable to the observer-flash distance, a more realistic assumption will be made in this section: filaments are short compared to the observer-flash distance $d$. This relative shortness indicates that it is relatively unlikely that the spherical tangent point will occur on the filament fragment, and that it is relatively probable that only a single real spot will ever exist on the filament fragment and move from one end to the other. Similarly, it is unlikely that the ellipsoidal tangent direction will point toward any specific short filament fragment, and that it is relatively probable that only a single perceived spot will be seen moving from one end of the filament section to the other. 

Goals of measuring perceived spots on finite filaments include determining the distances to the filament and the flash and finding the three-dimensional orientation of the filament fragment. Toward these ends, in general, noting a finite filament's angular distance from the flash direction does not constrain the observer-filament distance. The filament segment could exist at any location on a pair of lines starting at the observer and going through the ends of the segment. Noting the angular length of the segment would not help, in general, as the segment could be short and near the observer or long and far from the observer.

Furthermore, the apparent angular orientation of the segment does not imply a unique physical orientation for the segment, even if the distance to the finite filament is known. For example, an observed short segment oriented nearly perpendicular to the observer's line of sight would appear similar to a much longer segment oriented nearly parallel to the observer's line of sight. It would not even be clear which end of the segment is nearer the observer!

However additional information beyond the angular locations and times of perceived spots may be available that can resolve distance and orientation degeneracies.  Other commonly measured pieces of information include the distance to the flash $d$. are the time of the flash, $t_f$, from which relative times $\tau$ of angular echo observations can be obtained. Given a single $d$ and $\tau$, a unique time delay ellipsoid (TDE) -- a two dimensional surface -- is defined inside of which both the flash and observer are foci. A line from the observer through the $\theta$ of the filament fragment intersects this TDE two-dimensional surface exactly once, therefore isolating a single point in three-dimensional space. Stated mathematically, $\tau$, $\theta$, and $d$ can all be input into Eq. (\ref{PointDistance}) to give a unique $r$. Furthermore, if $\theta$ and $\tau$ are measured for the same perceived spot observed at another time, this second filament segment also becomes uniquely identified in space. Connecting two points in space yields an orientation of the line connection them, which therefore completely orients the linear filament in space as well. 

In more restrictive circumstances, suppose the observer makes one perceived spot observation and so knows $\theta_1$ and $\tau_1$, but not the distance $d$ to the flash. One cannot use Eq. (\ref{PointDistance}) to determine $r_1$, the distance to this perceived spot during a first observation, because, with $d$ unknown, there is one equation with two unknowns. If the observer assumes that the perceived spot is physically near the flash, then $r \sim d$, one now has one equation and one unknown, and Eq. (\ref{PointDistance}) can be used to estimate them both simultaneously. 

But now let's say that the same perceived spot was observed for a second time on the same linear filament segment, so that $\theta_2$ and $\tau_2$ become known. Using Eq. (\ref{PointDistance}) for each observation gives two equations -- but now there are three unknowns: $d$, $r_1$, and $r_2$. If one assumes the filament is short so that $r_1 \sim r_2$, then the situation reduces to two equations with two unknowns and the distance $d$ and distances $r_1 \sim r_2$ can be solved. However, assuming that $r_1 \sim r_2$ sacrifices the ability to orient the filament.

Let's now say that a third observation of this perceived spot is added so that $\theta_3$ and $\tau_3$ become known. Following the previous logic, Eq. (\ref{PointDistance}) can be used for each observation to create three equations -- but now there are four unknowns: $d$, $r_1$, $r_2$, and $r_3$. But the ability to orient the filament is different. Assuming a linear filament, a fourth equation exists stating that $r_1$, $r_2$, and $r_3$ all fall on the same line. Therefore given this third observation and corresponding fourth equation, it again becomes possible not only to find the distance to the flash and the filament but to orient a linear filament as well. 

If the distance to the flash is known but the time of the flash is not, then this time $t_f$ can be solved for using the same logic of the proceeding paragraphs. Another piece of discerning information when attempting to recover the three-dimensional orientation of the filament is a polarization measurement of the light scattered by the dust in the filament, as it should be correlated with deflection angle.

\section{Intervening filaments}

Light echoes may be particularly noticeable when they occur on a filament passing directly between the observer and the flash. Although this coincidental positioning may appear unlikely, it is actually a one-dimensional subset of a directly intervening two-dimensional sheet of dust, a case that has been observed many times as expanding flash-centered rings (for a recent review see, for example \citet{2012PASA...29..466R}). Given the correct time ordering, it is possible to connect echoes on filaments more generally to echoes on two-dimensional sheets.

It will be assumed in this section that the length of the filament section is very small when compared to $d$, the distance to the flash. It will be further assumed that the filament occurs all at a single distance $h$ from the observer. Therefore, the main goal of the analysis of this section is not to find the filament fragment's orientation, but rather, given measurements of $\theta$ and $\tau$, its distance $h$, and the distance to the flash $d$. 

Given these goals and approximations, a simple simplifying assumption will be made in this section: that that the linear filament's length is oriented perpendicular to the flash - observer axis. The spaghetti-thin geometry of practically-observed TDEs make this clear as any filament that crosses into a TDE likely crosses out at very nearly the same distance from the observer. Therefore, in this section, $y$ will refer to distance perpendicular to the flash - observer axis rather than along the filament. 

Additionally, the perigee direction will coincide with the flash direction. Effectively, the ellipsoidal tangent direction will also be in the direction of the flash, so that the perceived spot pair creation event will be seen directly toward the flash and both perceived spots will appear to angularly move directly away from the flash. 

How fast do the perceived spots appear to move? The described geometry as assumed, is depicted in Figure \ref{fig:TDE}. The variable $s$ here will be taken to parameterize the distance between the flash and the filament along the observer's line of sight. Therefore $d = h + s$. The distance from the flash to position $y$ on the filament is $c t_y = \sqrt{ y^2 + s^2}$. Note that here, $y = 0$ denotes the point on the filament that lies directly on the line connecting the flash to the observer. The distance from observer to position $y$ on the filament is $ c t_s = \sqrt{y^2 + h^2}$, where $t_s$ is the time it takes for light to go from position $y$ to the observer. Therefore the total distance that light flies from the flash to location $y$ on the filament to the observer is $c t = c t_y + c t_s = \sqrt{y^2 + s^2} + \sqrt{y^2 + h^2}$. To compute the speed of the perceived spots across the filament, one can take the $y$ derivative of both sides so that
\begin{equation} \label{SpeedGeneral}
 v = { c \over{{ y \over \sqrt{y^2 + s^2} } + { y \over \sqrt{y^2 + h^2} }} }.
\end{equation}

Since $h = d - s$, one parameter, $s/d\in[0,1]$, can be used to describe the position of the flash, the linear filament and the observer. As a practical limit, $y < 0.01 d$ in the following discussion.

\begin{figure}
\begin{center}
\includegraphics[width=0.7\columnwidth]{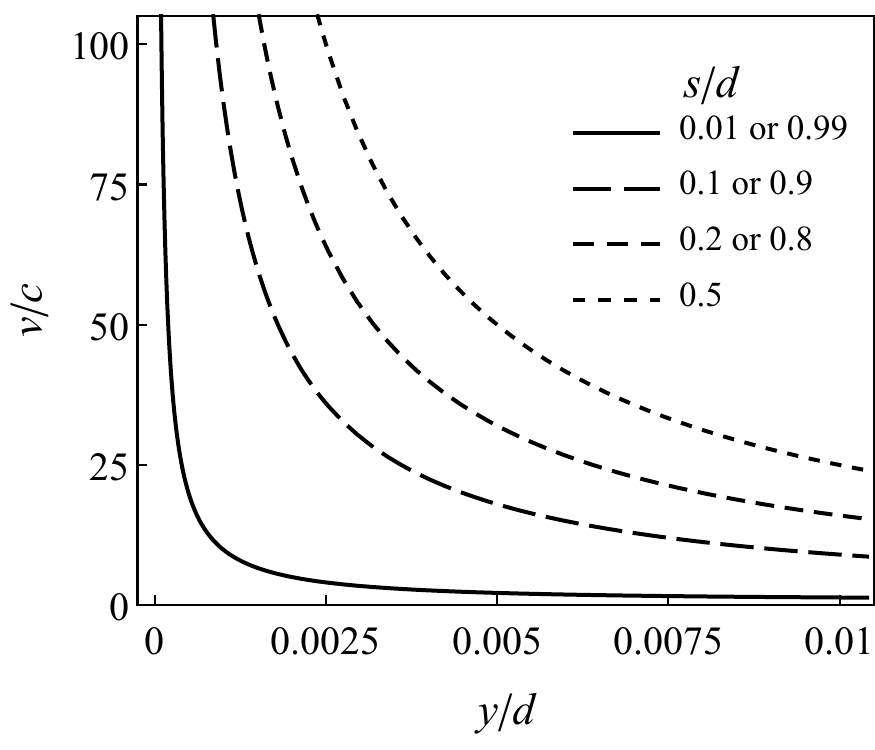}
\caption{{\label{fig:PerceivedVelocity}: The perceived speeds of the illumination fronts on the filament.%
}}
\end{center}
\end{figure}

Eq. (\ref{SpeedGeneral}) shows that, if $s$ is replaced by $(d-s)$, $v$ is the same. For example, given a value for $y$, $v$ is the same when $s/d=0.1$ and when $s/d=0.9$. This is a kind of symmetry. A plot of the perceived speeds of the perceived spots is given in Figure \ref{fig:PerceivedVelocity}. Eq. (\ref{SpeedGeneral}) can become markedly simpler if any of the approximations in the following subsections are invoked.

Writing Eq. (\ref{SpeedGeneral}) in terms of observables $\theta$ and ${\dot \theta}$, and considering that the perceived spots are always visible near the flash direction, then effectively $\sin \theta \sim \theta = y / h$ and ${\dot \theta} = v / h$. Then Eq. (\ref{SpeedGeneral}) can be written
\begin{equation} \label{AngularSpeedGeneral}
 {\dot \theta} = { c/h \over { \theta \over \sqrt{\theta^2 + (s/h)^2}} 
               +             { \theta \over \sqrt{\theta^2 + 1} } } .
\end{equation}
Explicitely writing $\theta^2 \ll 1$ and rearranging terms yields
\begin{equation} \label{ThetaThetaDot}
\theta ({\dot \theta} h / c) =   { \sqrt{\theta^2 + (s/h)^2} \over
                                  (1 + \sqrt{\theta^2 + (s/h)^2} ) },
\end{equation} 
where ${\dot \theta}$, $h$, and $c$ are grouped together to delineate the perceived spot speed in terms of the speed of light.

\subsection{At the Flash}

The special case where the scattering filament is right at the flash will first be considered. Here the line of the scatterer will be considered perpendicular to the line of sight. Mathematically, this case occurs when  $s = 0$. It will also be assumed that $y \ll d$. Then Eq. (\ref{SpeedGeneral}) becomes
\begin{equation} \label{SpeedAtFlash}
 v =  {c \over {1  + { y \over \sqrt{y^2+d^2} }}}\approx (1-{y \over d} )c\sim c,
\end{equation}

Therefore, the speed of the illumination front on the perpendicular linear filament is very close to $c$. This makes sense as that is the speed that light climbs the filament. Given that the angular separation of the perceived spot and the flash is $\theta= y / h = y / d$, then the angular speed of the illumination front as seen by the observer would be ${\dot \theta} = v / d = c / d$. Since this is constant, ${\ddot \theta} = 0$, meaning that the observer sees the illumination front move out from the flash at constant angular speed. Since this is the only distance far from the observer where this occurs, an expanding light-echo ring or sphere with constant angular speed must occur at the flash. In this special case, it is straightforward to compute the distance to the flash as $d = h = c / {\dot \theta}$. 

Were a flash embedded in an extended but optically thin scattering medium, then the perceived outermost spherical illumination front might yield the distance to the flash in this way. Note, however, that this is {\it not} thought to be the case in V838 Mon, where the flash is thought to be reflected from surrounding spherical shells of fixed radii \citep{2003Natur.422..405B}. An independent method for determining that the filament is perpendicular to the flash is by the polarization of the light received from the filament \citep{1994ApJ...433...19S}.

\subsection{Near the Flash}

Another case of interest occurs when the scattering sheet is known to be near, but not at, the flash. An example would be a scattering dust sheet in the same galaxy as a distant supernova. Specifically, isolating a line from that sheet that passes between the flash and observer defines the scenario considered in this subsection. Mathematically, it will be assumed that the scattering distance $s$ is much closer to the flash than the observer: $s \ll d$, but at the same time much larger than distances on filament: $s \gg y$. Therefore $\theta \sim y/h \ll s/h$, and $s/h \ll 1$. Then Eq. (\ref{ThetaThetaDot}) becomes
\begin{equation}\label{SpeedNearFlash}
{ \theta (h {\dot \theta} /  c) } = {s \over h} .
\end{equation}
Given that $d t = d \tau$ and that $\theta = 0$ when $\tau = 0$, it is straightforward to integrate Eq. (\ref{SpeedNearFlash}) so that 
\begin{equation} 
\theta = {\sqrt{ 2 s c \tau} \over h} .
\end{equation}
This can be written in terms of $\theta$ and $\tau$ such that
\begin{equation} \label{ThetaSH}
{ \theta \over \tau^{1/2} } = \left( {2 s c \over h^2} \right)^{1/2}. 
\end{equation}
Given specific measurements of $\theta$ and $\tau$, there are infinitely many $s$ and $h$ pairs that solve Eq. (\ref{ThetaSH}). Therefore even meticulous tracking of $\theta$ with $\tau$ will not resolve both $s$ and $h$. However, once either $s$ or $h$ is known, the other can be solved for directly.

\subsection{Between the Flash and the Observer but Far from Each}

Here it will be assumed that $y \ll s$ and $y \ll h$, meaning that the scattering filament is in the middle between the flash and the observer, but close to neither. In these cases, for example, $s/d$ might be between the values of 0.1 to 0.9. Then Eq. (\ref{SpeedGeneral}) becomes
\begin{equation} \label{SpeedMiddle}
 v\approx {c \over {{y\over s}+{y\over h}}} = {h s \over yd}c .
\end{equation}

A perceived spot pair creation event will appear at $y = 0$ when $\tau = 0$ and after some time $\tau$ both perceived spots will appear at the angular distance $\theta = y / h$ from the flash direction. 

As with the real spots, Eq. (\ref{SpeedMiddle}) shows that the speed for the perceived spots also starts at $y = 0$ with a formally infinite value and decreases as it gets further from the flash. The perceived angular speed of the perceived spot on the filament -- and hence the perceived speed of the expanding ring -- is just ${\dot \theta} = v / h$ so that ${\dot \theta} = s c / (y d)$ and
\begin{equation} \label{AngularSpeed}
 \theta ({h \dot \theta / c}) = { s \over d} .
\end{equation}
Note that Eq. (\ref{SpeedNearFlash}) is a special case of Eq. (\ref{AngularSpeed}) when $h = d$.  Given that $s = d - h$, Eq. (\ref{AngularSpeed}) can be solved for the distance to the scatterer as 
\begin{equation}
h = { d c \over d \theta {\dot \theta} + c } .
\end{equation}
Given that $d t = d \tau$ and that $\theta = 0$ when $\tau = 0$, it is straightforward to integrate Eq. (\ref{AngularSpeed}) so that 
\begin{equation} \label{MiddleH}
h = { 2 c \tau d \over \theta^2 d + 2 c \tau} ,
\end{equation}
which is a special case of Eq. (\ref{PointDistance}). This can be rewritten in terms of $\theta$ and $\tau$ such that
\begin{equation} \label{ThetaTau}
{ \theta \over \tau^{1/2} } = \left( { 2 c (d - h) \over d h } \right)^{1/2} .
\end{equation}
Given specific measurements of $\theta$ and $\tau$, there are infinitely many $d$ and $h$ pairs that solve Eq. (\ref{ThetaTau}). Therefore, even meticulous tracking of $\theta$ with $\tau$ will not resolve both $d$ and $h$. However, once either $d$ or $h$ is known, the other can be solved for directly.

\subsection{Near the Observer}

An example of a light scattering plane near the observer occurs when the X-rays from a cosmologically distant gamma-ray burst become scattered by a plane of dust in our Milky Way Galaxy (see, for example, \citet{2004ApJ...605L.101W}). Specifically, isolating a line from that plane that passes between the flash and observer defines the scenario considered in this subsection. Following logic similar to when the scattering filament was near the flash, when the filament is near the observer one finds the perceived spot speed becomes 
\begin{equation} \label{NearSpeed}
{dy \over dt} = { \sqrt{y^2 + h^2} \over y } c \sim {h c \over y} .
\end{equation}
Written in terms of angular observables it is found that $\theta (h {\dot \theta} /c) = 1$ which has the solution 
\begin{equation} \label{NearAngle}
\theta = \sqrt{ 2 c \tau \over h} ,
\end{equation}
where it was demanded that $\theta = 0$ when $\tau = 0$. Note that $d$, the distance to the flash, does not appear and so cannot be recovered for the approximations given. However, the distance to the scatterer $h$ can be determined by a single instance of measuring $\theta$ at a known $\tau$ with 
\begin{equation}
h = { 2 c \tau \over \theta^2 } .
\end{equation}

\subsection{At the Observer}

In analogy with the above considered case where the scattering filament occurred right at the flash, for further didactic purposes, the case where the scattering filament is right at the observer, perpendicular to the flash, will be considered. Mathematically, this case occurs when $h = 0$ and so $s = d$. Because the filament intersects the observer, the angular locations and speeds to "perceived" spots or rings are not actually visible to the observer because they are purely radial to the observer's position. However, inspection of Eq. (\ref{NearAngle}) indicates that both $\theta$ and ${\dot \theta}$ diverge as $h$ goes to zero. Together, these divergences indicate that when the spherical shell of light expanding from the flash reaches the observer, it is essentially a plane wave with no curvature.

\section{Brightness}

In general, the intrinsic brightness of a real spot, created by reflection, will involve the intrinsic brightness of the flash over time, the opacity of the light paths between the flash and the filament and between the filament and the observer, the thickness of the filament as a function of distance along its length, the angular efficiency of light scattering, and the wavelength dependence of all of these. For a real spot created by heating, the intrinsic brightness will also involve the composition and temperature of the gas or dust \cite{2008ApJ...685..976D}, while for a real spot created by ionization, the intrinsic brightness will involve the ionization potential, density, and ionization fraction of the illuminated elements.

Furthermore, the apparent brightness of a perceived spot will also involve line-of-sight integrals over many of these variables as well as the temporal, angular, and energy resolutions of the observing instruments. Therefore, besides some general considerations, the real and perceived brightness of the illuminated spots involves a tremendous amount of analyses and will be considered beyond the scope of this geometric treatment. The most important attribute of the brightness of a perceived spot is that it be different than the background brightness of the unilluminated filament. 

Even so, quite generally, assuming that attributes of the scattering media remain nearly constant, it is expected to lowest order that perceived spots will maintain at an approximately constant brightness, given that the distance that perceived spots move is small compared to other angular distances in the scenario. Past that, to the next order, perceived spots below the angular and temporal resolution of the detector are expected to appear brightest when they are moving the fastest because their instantaneous surface brightness is expected to be constant. Therefore high relative brightness should correspond to geometries nearest perceived pair creation and annihilation events, as that is where perceived spot speeds are the highest. When two spots from a pair are angularly far from each other as well as their perceived creation or annihilation locations, they may appear to be of significantly different brightness, and their brightness ratio may carry information about the relative orientations of the filament and the flash.

\section{Examples}

In this section two hypothetical example systems are considered in astronomical settings. To be clear, no measured light echo detections are being claimed from these systems. Rather these systems are considered here as examples primarily because they feature a bright source that is illuminating a nearly angularly straight filamentary segment of dust. The purpose of these examples is to demonstrate more concretely how distance and orientation information might be recovered, were the illuminating source of a filament to undergo variability that could later been seen with perceived spots on that filament. 

The filament segments considered here appear, in images, to be nearly linear. However, basic photography records only angular information, so that it is possible that either or both of these example filaments are really curved in space but along the perceived angular line of the filament. Nevertheless, for didactic purposes, it is considered here that the filaments are truly linear in three-dimensional space.

\subsection{Merope}

The bright star Merope (23 Tau, $m_v \sim 4$) in the Pleiades is observed to be surrounded by the bright reflection nebula IC 349 that contains numerous filaments that appear, angularly, nearly straight. The system was first studied over a century ago by \citet{1891AN....127..135B}, and more recently by others including \citet{Herbig_2001}. In particular, a single one of these filaments situated about 5 arcminutes away -- and stretching about 5 arcminutes -- will be considered. This filament, if extended linearly and angularly, would nearly intersect the direction of Merope, and so is approximately collinear, angularly, with Merope. The star Merope itself is about 118 pc distant \citep{1999A&A...345..471R}. Were the filament flat on the sky and all at the same distance as Merope, then the end nearest to Merope would also be as distant from Earth as Merope, and would be separated from Merope by about one light year. 

Even given that the filament is truly linear in three-dimensional space, there remains, unfortunately, an infinite number of three-dimensional spatial orientations that would result in its observed two-dimensional angular orientation. For example, a long filament segment situated nearly along the line of sight, or a short filament segment situated nearly perpendicular to the line of sight could appear, from Earth, similar. Furthermore, angular observations of the filament cannot even resolve the ambiguity of which end of the filament is physically closest to Merope.

Fortunately, this orientation ambiguity can be resolved, in theory, first by noting the distance to Merope and the time of a flash, and then by observing the corresponding pattern of brightness changes on the filament in response to that flash. Stated differently, given $d$ and $t_f$, each linear filament orientation would yield a different perceived pattern of illumination change from a flash from Merope. 

\begin{figure}
\begin{center}
\includegraphics[width=0.7\columnwidth]{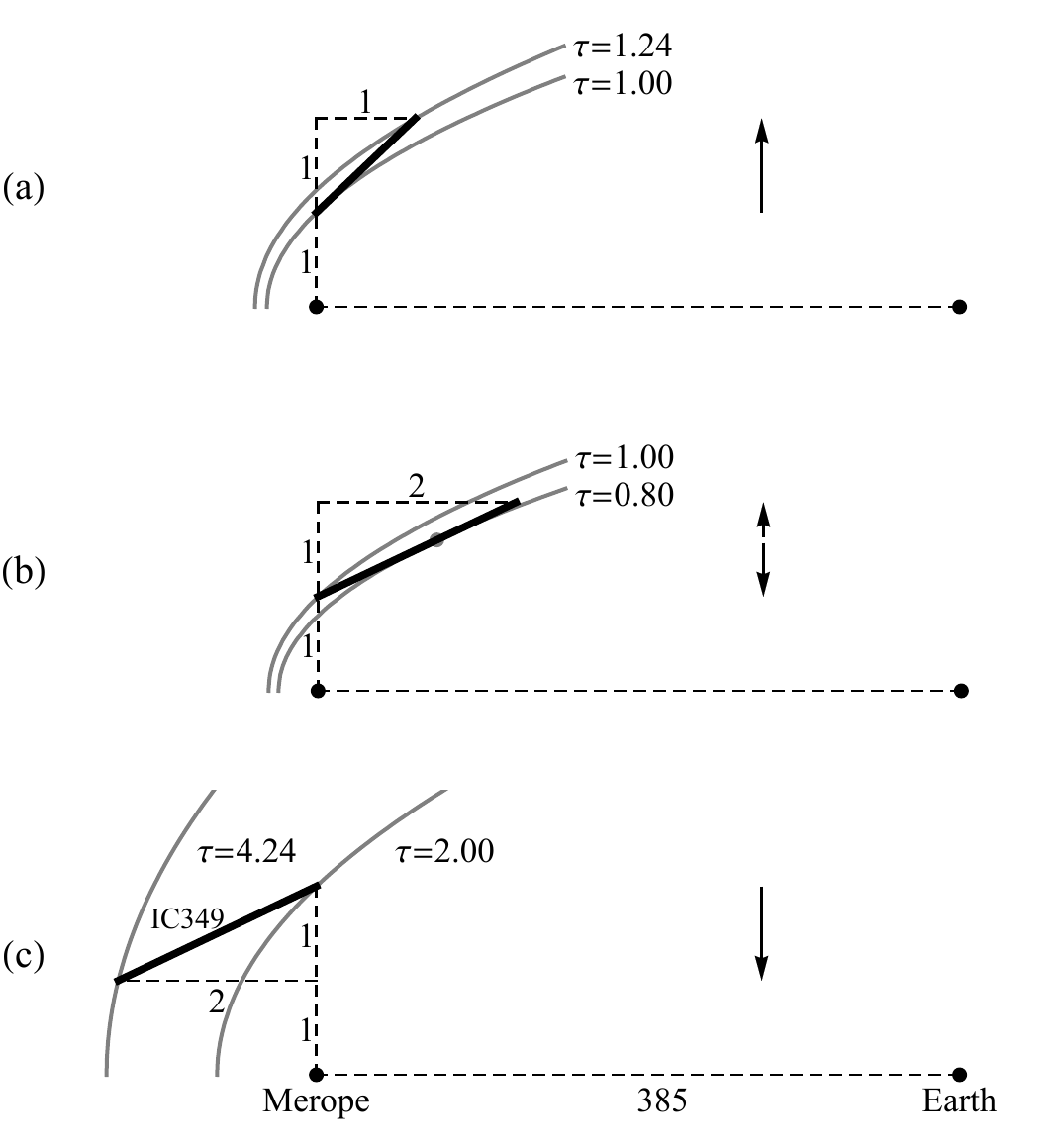}
\caption{{\label{fig:Merope}: Hypothetical geometries of a linear filament in IC 349 near the star Merope. The arrows show the motion of the perceived light spots. Sections of two relevant time delay ellipsoids (TDEs) are shown.%
}}
\end{center}
\end{figure}

To demonstrate this technique more concretely, three different distance and orientation geometries for the specified linear filament near Merope will be assumed, all corresponding to what is seen from Earth angularly. In the first example, the angular end closest to Merope will be considered to lie precisely at the same distance of Merope (118 pc, 385 ly) and 1 light year on one side of Merope. The other end, angularly farthest from Merope, will be assumed to lie 1 light year closer to Earth and 2 light years to the same side of  Merope. Therefore, the actual physical length of the filament would be about 1.41 light years. The projected length of this filament in Earth's sky is 1 light year. This geometry is depicted in Figure \ref{fig:Merope} (a).

Consider that Merope emits a short duration flash at time $t_f = 0$. First the real illumination pattern on the example filament will be described. The flash creates an annular spherical shell of light that moves out from Merope. This flash will create a real spot of illumination that moves along the example filament. Were the filament of the same orientation but infinite extent, the flash would first illuminate the filament at the spherical tangent point. After that, two real spots of illumination would move in opposite directions along the filament. Given the orientation and limited extent of this example fragment, however, only one of these real spots will occur. As the spherical shell of light expands, the nearest end of the filament to Merope will be illuminated first, which would occur 1 year after Merope's flash. A real spot of illumination will move along the filament until it reaches the other end of the filament fragment. As this end is 2.24 light years from Merope, this filament end will be illuminated 2.24 years after the flash which will be 1.24 years after the first filament end was illuminated. Note that the real spot moved the 1.41 light years in 1.24 years, which corresponds to about 1.14 times speed of light. The superluminal real spot speed is consistent with {\it all} real flash illumination fronts being superluminal \cite{2016PhyEd..51d3005N} and does not mean that any massive object was moving superluminally.

Next, the perceived illumination pattern on the filament will be found. On Earth, the flash from Merope is seen first, 385 years after the flash actually occurred. This is defined as $\tau = 0$. The light echo from the linear filament is seen only after light from the flash reaches the filament and then travels to the Earth. In the example, the nearest end of the filament to the Merope is actually illuminated one year after the flash and therefore, since it is at the same distance from Earth, also appears to be illuminated, on Earth, one year after the flash. The other end of the filament is 2.24 light years from Merope, and so becomes illuminated 2.24 years after the flash. However, since this end is one light year closer to Earth, light takes one year less to reach Earth from this end than from the other end. Therefore, this end is seen to be illuminated at the Earth 1.24 years after the flash was recorded. The TDEs for both $\tau = 1.00$ years and $\tau = 1.24$ years are also shown in \ref{fig:Merope} (a). The perceived illuminated spot on the filament appears to move away from Merope, appearing to go from the end nearest Merope to the end furthest from Merope in 0.24 years. Note that were the filament considered to actually have its projected length of 1 light year, the perceived spot would appear to have a speed of 4.17 c.

In a second example using the same star, a different three-dimensional orientation for the filament will be assumed that displays the same apparent angular orientation and extent on the sky. Here it will be assumed that the filament is oriented in such a way that a perceived pair-creation illumination front appears on the filament from the vantage point of the Earth. In other words, the ellipsoidal tangent direction will occur in a direction along the length of the filament segment. As such, one perceived spot on the filament appears to move angularly toward the flash, while the other perceived spot moves angularly away. Here, the angularly farthest end of the filament lies 2 light years closer to Earth instead of 1 light year, as shown in Figure \ref{fig:Merope} (b). The filament end closest to Merope is seen on Earth 1 year after the flash from Merope. Light takes 2.83 years to reach the other end of the filament, but since this end is 2 light years closer to Earth, the end is seen illuminated only 0.83 years after the flash. However, an observer on Earth will first see a perceived spot pair event from a position which is 1.2 light years closer to Earth on the filament than Merope. That this perceived pair event is seen first is demonstrated by the fact that it only takes 0.80 years, after the flash, for light from this direction to reach Earth. It is interesting to note that for this filament orientation, no real spot pair event occurs on the filament at all. Even though a perceived spot pair event occurs, there is only one real spot from the flash on the filament segment at any time. 

In a third and final example involving Merope and the same angular filament, the filament will be considered to be oriented in such a way that a single perceived spot of illumination appears, as viewed from Earth, and this single perceived spot moves angularly {\it toward} the flash, not away from it. In this example geometry, the filament in Figure \ref{fig:Merope} (b) will be slid back and placed so that its furthest end lies 2 light years away from Earth, as shown in Figure \ref{fig:Merope} (c). A single perceived spot appears because the ellipsoidal tangent direction again does not occur in the direction of this example filament. Here, light takes 2.24 years to reach the filament end that is angularly closest to Merope, but since this end is 2 light years further from Earth, this end is seen from Earth to be illuminated 4.24 years after the flash. However, the other end of the filament, the one angularly furthest from Merope, is seen on Earth only 2 years after the flash from Merope. It is interesting to note that for this filament orientation, a real spot pair event does occur on the filament fragment, even though only a single perceived filament spot is the most ever observed from Earth.

In reality, Merope has never been seen to undergo a single bright flash. However, Merope, classified as a Beta Cepheid variable, has been detected undergoing variability at about 0.01 magnitudes over 0.5 days \citep{1971AJ.....76.1058C}. Were this variability to deviate from strict periodicity, it would be clearly possibly to determine the three dimensional geometry of the filament using the logic of this section. However, even given an exact periodicity, it may also be theoretically possible to resolve the spacial geometry of the filament by angularly and temporally resolving illumination fronts on the filament.

\subsection{IC 2118: The Witch Head Nebula}

A second example astronomical system is IC 2118, the Witch Head Nebula. Although the reflection nebula has details reminiscent of the head of a witch, the nebula's basic angular extent is, to a good visual approximation, linear. IC 2118 spans about 5 degrees in length and has an estimated distance of about 210 pc \citep{2001PASJ...53.1063K}. Combined, these translate into a projected length on the sky for IC 2118 of about 60 light years.

The dust in IC 2118 would appear dark were it not for the light of the bright nearby star Rigel (Beta Orionis, $m_v = 0.13$), which appears about 2 degrees away. Because of Rigel, the nebula glows distinctly blue. For the didactic purposes of this example, it will be assumed that IC 2118 is completely flat on the plane of the sky, and that Rigel and all of IC 2118 are the same distance from Earth. Of course, many other three-dimensional orientations of the IC 2118 filament could result in the angular orientation seen, and it is not presently known which orientation IC 2118 really has. Given this assumed orientation, though, many attributes of perceived light echoes will occur toward the directions of the real light echoes. Inspection of an image of IC 2118 and Rigel \citep{2008APOD...OCT...31} indicates that the closest angular approach of IC 2118 to Rigel occurs about 1/3 of the way along the filament. In this example, this direction corresponds to the spherical tangent point and is seen from Earth in the ellipsoidal tangent direction.

Given the example geometry, a bright flash from Rigel would be followed about 24 years later by the first light echo from IC 2118 at the spherical tangent point, which here is seen from Earth in the ellipsoidal tangent direction. The observer would then see two perceived spots move out towards opposite ends of the filament. The perceived spots would start with formally infinite angular speed but quickly slow. Each perceived spot would have the same angular speed as its twin. The perceived spot closest to a filament endpoint would reach this endpoint first, about 7 years after the perceived spot pair creation event, and disappear. About 15 years after that, the other perceived spot would reach its filament endpoint and also disappear. 

Although Rigel is not known to generate bright flashes, the star is known to be an irregular variable \citep{1933ApJ....77..226S} \citep{2010HiA....15..359G}. In principle, were Rigel's variations irregular enough, bright enough, and tracked long enough, they could be found in relative reflections from IC 2118. Measured time offsets would yield, in theory, the relative orientation of IC 2118 relative to Rigel. Then, were the distance to either IC 2118 or Rigel found with high accuracy, the distance to the other could be calibrated.

\section{Discussion and Conclusions}

Light echoes are well known phenomena seen in visible, infrared, and X-ray light. In this paper, the concept of a light echo is considered in detail for scattering from a simple linear filament. The analysis given is primarily geometric and does not rigorously address brightness. Much of the analysis allowed the filament to exist anywhere in space, not being confined near the flash. Given a flash, it was discussed how real spots of illumination move on a linear filament irrespective of an observer, including how real spots are created in pairs that move apart. Different from real spots, it was shown how perceived spots of illumination may appear to move to an observer. For conceptual background, an infinite filament was first considered for perceived spot motions and directions of interest, including apogee, perigee, and radiant directions defined. It was shown that a perceived spot pair event is the first echo illumination event that appears for a flash on an infinite filament, and that one of these perceived spots must move with an angular component {\it toward} the angular location of the flash, rather than away. Such angular motion has, to the best of our knowledge, never yet been seen nor even explicitly discussed in an astronomical setting. Details were derived of what types of observations -- and how many -- are needed, in theory, to completely orient a filament in space. It was noted that the observation of a perceived spot pair creation event gives more information than the observation of a single perceived spot, information that may be used to help resolve source and filament distance and geometry. For filaments intervening between the observer and the flash, it was shown what observational information is needed to recover the distance to the filament and the flash. 

Were a candidate perceived spot pair identified, its character may need to be discerned from the case of a single perceived spot illuminating a complex dust filament. Observed over a long enough time, the motions of spot pairs on linear filaments should not only be symmetric but predictable in location and speed, making this type of event uniquely identifiable. 

This formalism presented can also be applied to linear filaments that have perceived spots of illumination that do not result from direct scattering of light from a flash, but rather emission from heating from a flash \cite{1983ApJ...274..175D}, or even emission or absorption from ionization.

Generalizing to flash illumination of curved filaments, real and perceived spot pair events should also arise. Simple examples involving curved filaments, filament loops, and nearly straight filaments with segments of curvature should also show pair episodes. Furthermore, it is relatively straightforward to see that, in general, filaments with segments of curvature open away from the flash can exhibit real and perceived spot pair creation events, while curved filament segments open toward the flash can exhibit both real and perceived spot pair annihilation events. Filament curvature may act to change the expected locations and speeds of both real and perceived spots from that expected from a linear filament. Given the tremendous variety of curved filaments possible, quantifying illumination changes generally appears quite complex and is considered beyond the scope of the present work. More complex filament geometries, including curved geometries, will be examined in greater detail in a later work \citep{NemiroffInPrep2017}. A detailed example of how a flash illuminates a flat surface has been recently completed by \citet{Zhong:16}. 

Linear perceived spot motion holds promise to become practically observable in the foreseeable future, as large angular portions of the night sky are being observed with increasing frequency. For example, in 2008 the Pann-STARRS \cite{2004AN....325..636H} project began recurrent monitoring of much of the sky above Hawaii . In the near future, LSST will begin an even more ambitious sky monitoring campaign \cite{2009arXiv0912.0201L}. Additionally, the availability of increasingly inexpensive equipment to amateur observers has led to marked increase in repeated observations of wide areas of sky. Two remarkable examples include monitoring of Hubble's Variable Nebula for five months by Polakis (\url{http://www.pbase.com/polakis/image/163069532/original}), and a recent cumulative 212-hour exposure of the entire constellation of Orion by \citet{2015APOD...NOV...23}. Even with present technology, known filaments might be observed repeatedly with the objective of noting moving perceived illumination fronts or even perceived spot pair events.

~
\\
\linebreak \newline
\noindent{\bf ACKNOWLEDGEMENTS}

The authors acknowledge helpful comments from Chad Brisbois, Arlin Crotts, David Russell, Matipon Tangmatitham, and Teresa Wilson. We thank two anonymous referees for helpful and insightnful comments. QZ acknowledges support from the China Scholarship Council.

\bibliographystyle{mnras}
\bibliography{converted_to_latex}

\end{document}